# A Time Series Data Analysis of Indian Commercial Dynamism

**Samsul Alam**
Department of Business Administration
Daffodil International University, Ashulia, Savar, Dhaka

## Contents

In this report it is analyzed the focuses of the commercial dynamism of India, covering the fundamentals of growth rate, trade balance, coverage rate, openness rate, share of world indicators and then present each of them in detail.

## 1. Evolution of Trade (Goods and Services)

### 1.1 Average Growth Rate

The average growth rate is going to indicate the rate of growth that India has had in the study period.

The average rate of growth in exports of goods and services in India for the 14 (2000-2014) years is 8.52%, which indicates that every year India has presented a growth in exports to abroad. The average growth rate in imports over the period of study is 8.18% which also presents growth in imports from abroad. But if we compare this result with the growth rate of exports and imports we see that India has grown in exports that is, its growth rate is positive, which makes us think that India has been presenting a surplus in its current economy and dynamism concentrate on exporting side (See figure 1).

## 1.2 Trade Balance

Trade balance which is the difference between exports and imports of India is found during the study over the period of 14 years. From table 1, we can see that the trade balance of India made has positive figure for all economic years except 2012 where relevant data of 2000-2004 were not found in the dataset from where the study is done. In 2005 the trade balance was 19890000000.00 dollars showing positive figure that continued until 2011. In 2012 it had trade deficit as the trade balance was experiencing a negative value of 6404000000.00 dollars since the economy of India suffered economic instability that reduced gross domestic product (GDP) and the increasing buyer purchasing power and quality of life of people. Surprisingly in 2014 it grew in a huge amount of value of 80721000000.00 dollars, showing a surplus in its trade balance, which is an important change from the previous year. In that period India was overwhelmed by the economic sustainability for increasing GDP which allowed exporting a huge amount much greater than all the previous years and at the same time expensed lower amount than earlier 3 years (See figure 2).

## 1.3 Coverage Rate

The coverage rate is the ability of exports to finance imports. If the result is more than 100, it indicates that exports are able to fund imports; therefore India will enjoy a trade surplus. Contrary, it will suffer a trade deficit.

In table 1, you can see the result of the coverage rate of India of the years. It is observed that only in 2012; exports had inability to finance (98.58) all the imports showing trade deficit in the trade balance, since then instability was present. On the other hand, from 2005-2011 and 2012-2014 exports were able to finance all the amounts for imports. However, there was variation in every year but the growth rate was stable, matching a surplus in the trade balance where the highest surplus found in 2014 (119.93).

India has a higher growth for foreign trade. In many cases, India can be considered as the role model for countries those try to be experiencing trade surplus especially for Asian countries which are developing constantly (See figure 3).

## 2. Indicators of Trade Openness

### 2.1 Export Propensity

Export propensity is the ratio of exports to the GDP. It distinguishes the factors influencing whether or not subsidiaries are exporting. The table 2 shows the export propensity for various years. From the table, we see that the highest export propensity found is 25.23 (2013) for India where 31.55 (2008) for world and lowest is 18.53 (2005) for India and 26.63 (2009) for world and also found export propensity of other years are dynamic. Export propensity for India not bad comparing other developing countries.

### 2.2 Trade Openness Rate

The degree of trade openness is calculated as the ratio of total imports and exports to GDP, so that the higher the openness rate is, the higher the country or economic zone will depend more heavily on trade with the outside. The openness rate measures the degree of economic openness of a country. In other words, it is the ability of a country to market their goods and services with the rest of the world.

Figure 4 shows the percentages of openness of India. When the result tends to "0" it means that the foreign trade of the country being analyzed has little or no trade with the other countries. When the result tends to "100" it means that most of what it produces is exported and there is no contribution of domestic product to meet domestic purchaser needs.

India's total trade has variability over the last 14 years but this is ranged between 34.64% (2005)-48.81% (2012). In 2005, domestic products dedicated to the domestic market were 65.36% and in 2014 were 51.19% because of the increasing needs of the Indian population it tends to import more. Table 1 shows that the openness rate of India is varying, the result indicates that the country is partially dependent on foreign trade because it is one of the world largest countries that produce a variety of goods but these production can't fully meet the whole population demand. As no country does not sufficiently produce everything you need, India also does not produce all you need.

## 3. Indicators of Participation in World Trade

The next figure 5 illustrates that during the period 2005 - 2008, the exports and imports of India were rising rapidly with the time but in 2009 both sides dropped suddenly because of the economic instability. Again, after that period they started to rise and until 2012 it was continuing. After 2012 imports fell down and exports raised up that indicate a good economic surplus. In other words, India has earned weight in world trade (competitiveness), while participating in the world trade with a significant portion.

Concerning the total trade, India's main exports are mineral fuels and oils, natural or cultured pearls, vehicle; nuclear reactors, organic chemicals, pharmaceutical products, cereals, electrical machinery, cotton, iron & steel, apparel & clothing accessories, plastic, fish & crustaceans, aircraft, meat, textile articles, ships, and miscellaneous goods etc. On the other hand, India's main imports are crude petroleum, gold & silver, electronic goods, pearls & precious stones etc. The destinations of Indian exports are focused on United States of America (USA), United Arab Emirates (UAE), Singapore, China, Hong Kong, Saudi Arabia, Netherlands, United Kingdom (UK), Germany, Brazil etc. Again, the exporter countries to India are China, UAE, Saudi Arabia, USA, Switzerland, Iraq, Qatar, Kuwait, Germany, Indonesia etc.

However; in Figure 5 and 6; it can be seen that the global share of total trade of the country under study is relevant from the exporting side because during the 14 years shows that X>M. The following indicators serve to measure the dynamism and adaptation of a dynamic economic development and international trade for India.

### 3.1 Share of World Exports

The purpose of this indicator is to present the percentage of India's participation in world exports. Table 3 shows the percentage of India's share of exports of goods and services in world trade exports. It is assumed that the lowest share (1.02%) found in 2005 and the highest participation (2.06%) in 2014.

### 3.2 Share of World Imports

The purpose of this indicator is to know the percentage of imports that India makes in the total world imports. As shown in Table 3, the percentages in different years show the lowest participation (1.34) in 2005 and the highest participation (2.55) in 2012.

### 3.3 Share of World Total Trade

This indicator relates the imports and exports of India with world imports and exports of goods and services. India's participation for total share of world was highest (2.22) in 2012 and lowest (1.26) in 2005.

## Appendix

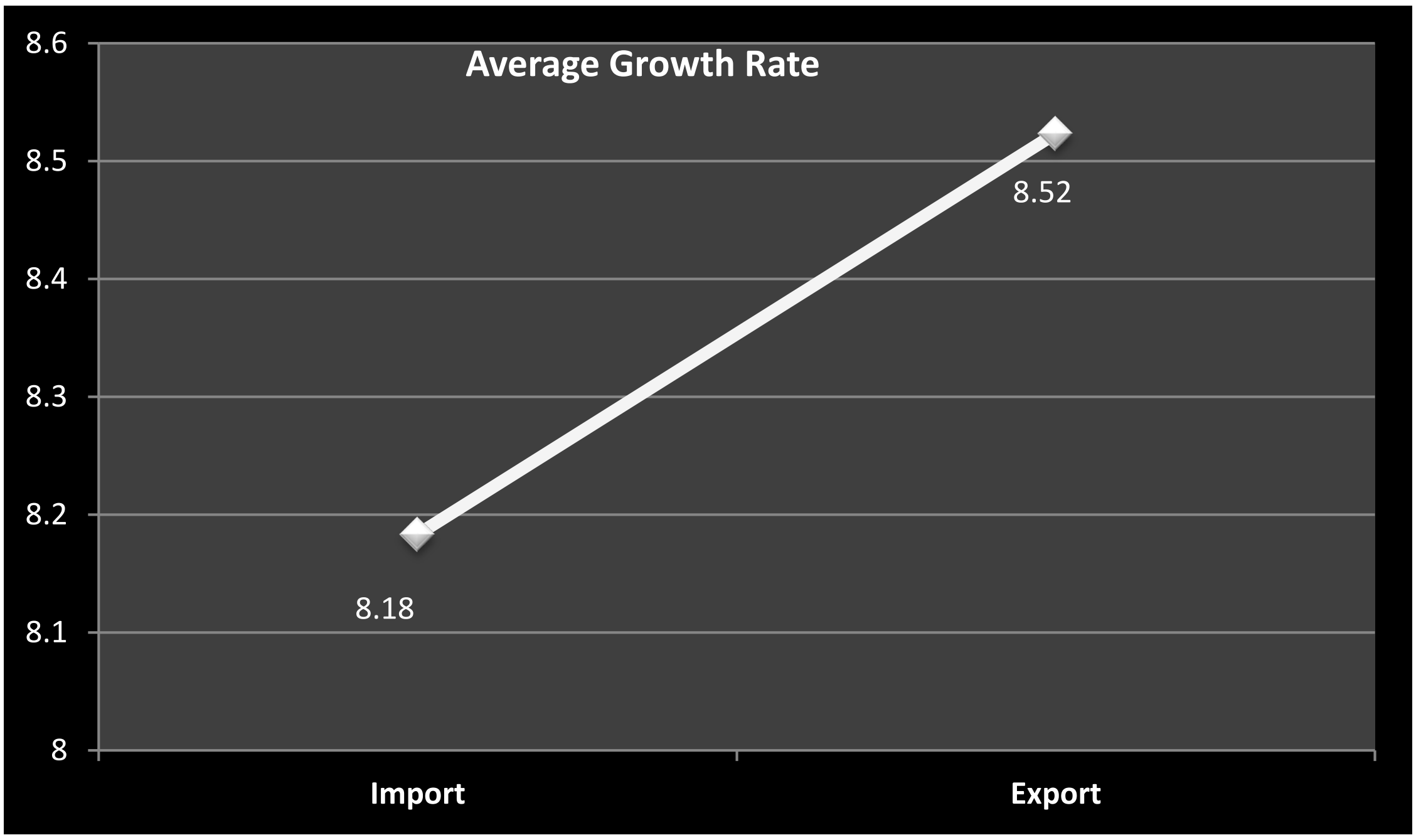

Figure 1: Average Growth Rate of India

Table 1. Year wise trade balance data of India with coverage and openness rate

| *Year* | *Export* | *Import* | *Trade Balance* | *Coverage Rate* | *Openness Rate* |
|---|---|---|---|---|---|
| *2000* | .. | .. | | | |
| *2001* | .. | .. | | | |
| *2002* | .. | .. | | | |
| *2003* | .. | .. | | | |
| *2004* | .. | .. | | | |
| *2005* | *154582000000.00* | *134692000000.00* | *19890000000.00* | *114.77* | *34.68* |
| *2006* | *193316000000.00* | *166572000000.00* | *26744000000.00* | *116.06* | *37.92* |
| *2007* | *240082000000.00* | *208611000000.00* | *31471000000.00* | *115.09* | *36.22* |

| *Year* | *Export* | *Import* | *Trade Balance* | *Coverage Rate* | *Openness Rate* |
|---|---|---|---|---|---|
| *2008* | *305119000000.00* | *291740000000.00* | *13379000000.00* | *104.59* | *48.76* |
| *2009* | *260847000000.00* | *247908000000.00* | *12939000000.00* | *105.22* | *37.26* |
| *2010* | *348035000000.00* | *324320000000.00* | *23715000000.00* | *107.31* | *39.35* |
| *2011* | *446375000000.00* | *428021000000.00* | *18354000000.00* | *104.29* | *47.63* |
| *2012* | *443845000000.00* | *450249000000.00* | *-6404000000.00* | *98.58* | *48.81* |
| *2013* | *467759000000.00* | *433760000000.00* | *33999000000.00* | *107.84* | *48.42* |
| *2014* | *485843000000.00* | *405122000000.00* | *80721000000.00* | *119.93* | *43.11* |

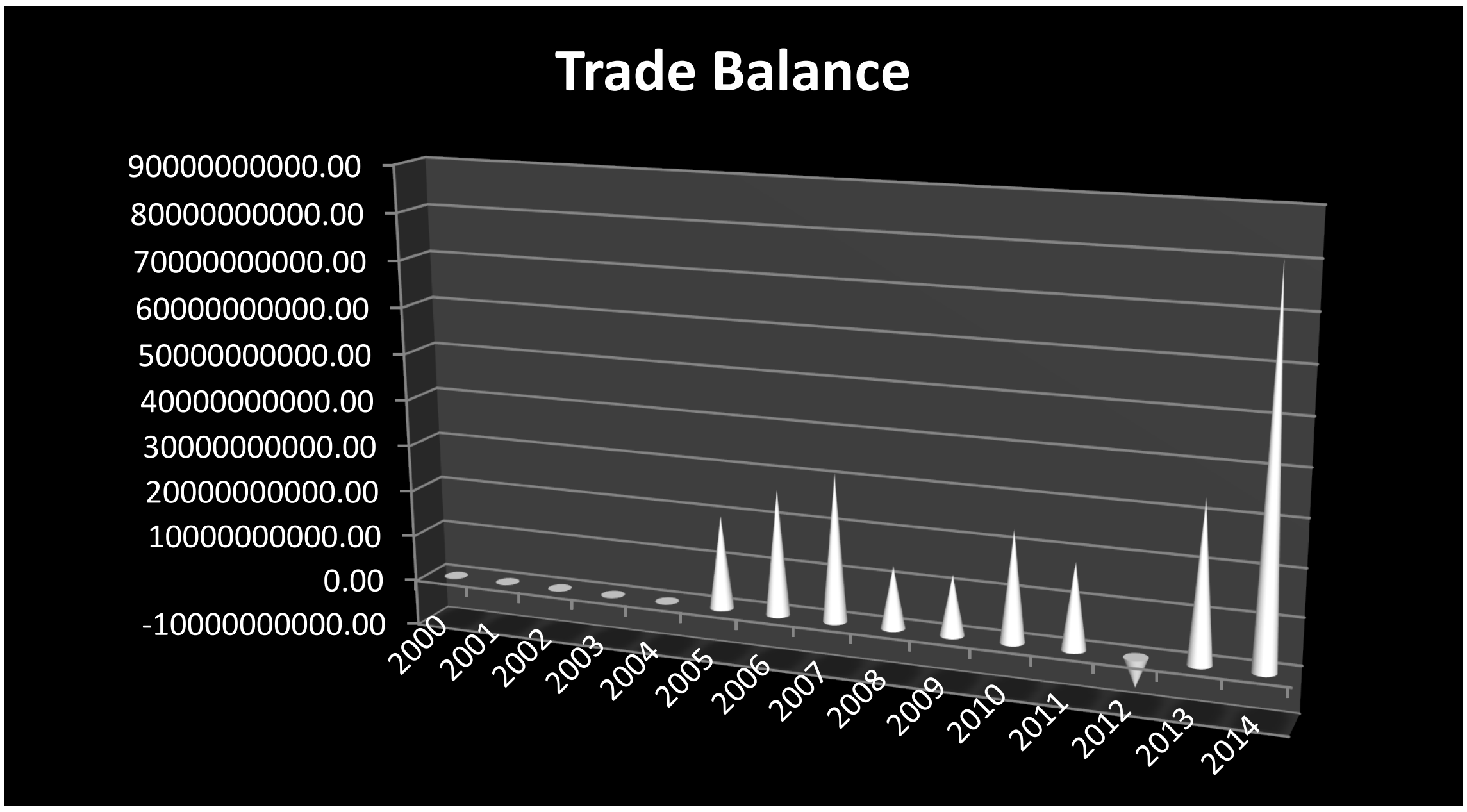

Figure 2: The trade balance of India

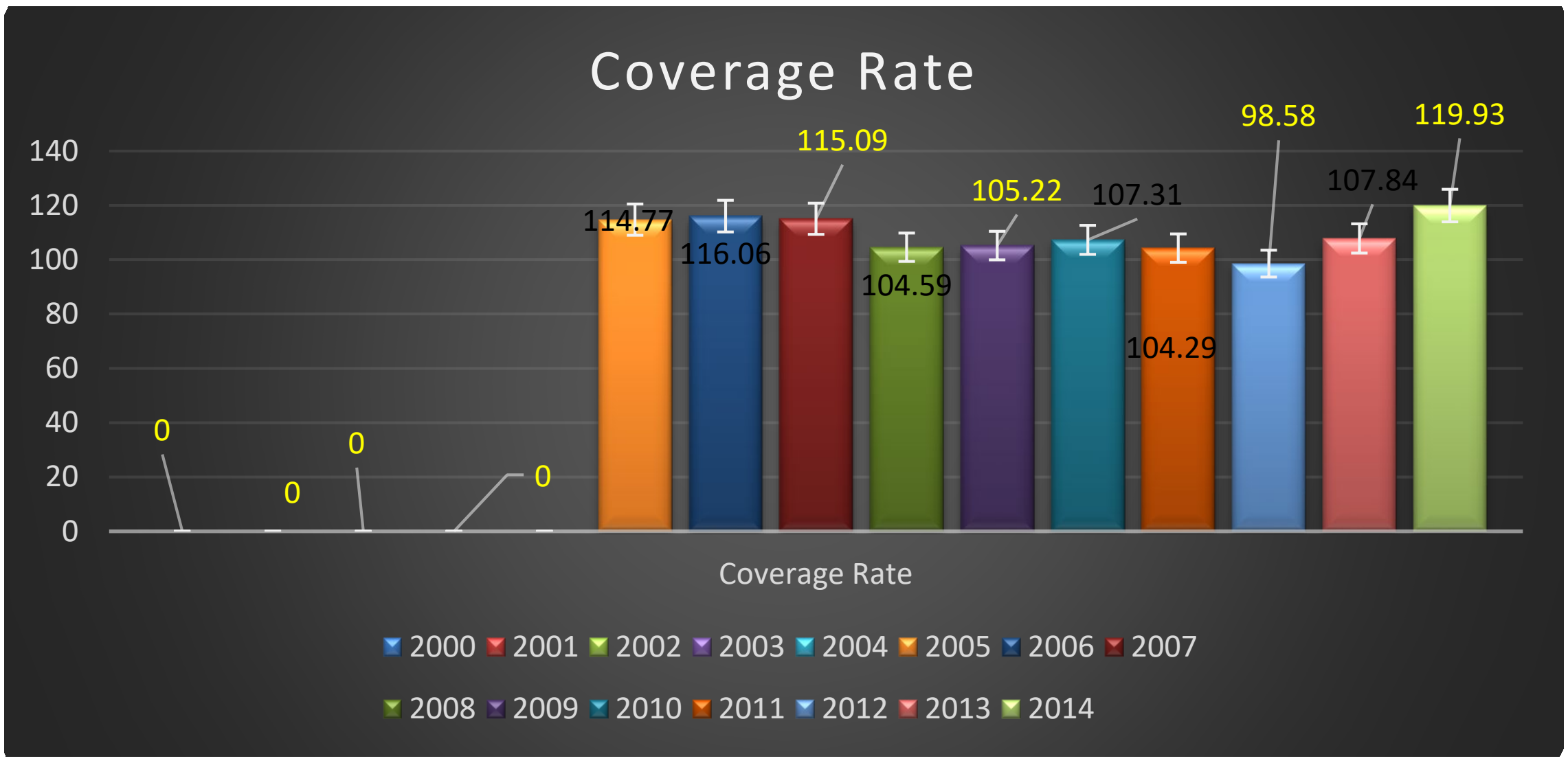

Figure 3: The coverage rate of India

Table 2. Year wise export propensity of India

| ***Year*** | ***Export*** | ***GDP*** | ***Export Propensity*** |
|---|---|---|---|
| *2000* | .. | 476609148165.17 | |
| *2001* | .. | 493954333981.34 | |
| *2002* | .. | 523968561883.92 | |
| *2003* | .. | 618356467437.03 | |
| *2004* | .. | 721585608183.52 | |
| *2005* | *154582000000.00* | 834215013605.93 | *18.53* |
| *2006* | *193316000000.00* | 949116769619.55 | *20.37* |
| *2007* | *240082000000.00* | 1238699170079.01 | *19.38* |
| *2008* | *305119000000.00* | 1224097069459.66 | *24.93* |
| *2009* | *260847000000.00* | 1365371474048.19 | *19.10* |
| *2010* | *348035000000.00* | 1708458876829.92 | *20.37* |
| *2011* | *446375000000.00* | 1835814449584.80 | *24.31* |
| *2012* | *443845000000.00* | 1831781515471.56 | *24.23* |
| *2013* | *467759000000.00* | 1861801615477.90 | *25.12* |
| *2014* | *485843000000.00* | 2066902397333.26 | *23.51* |

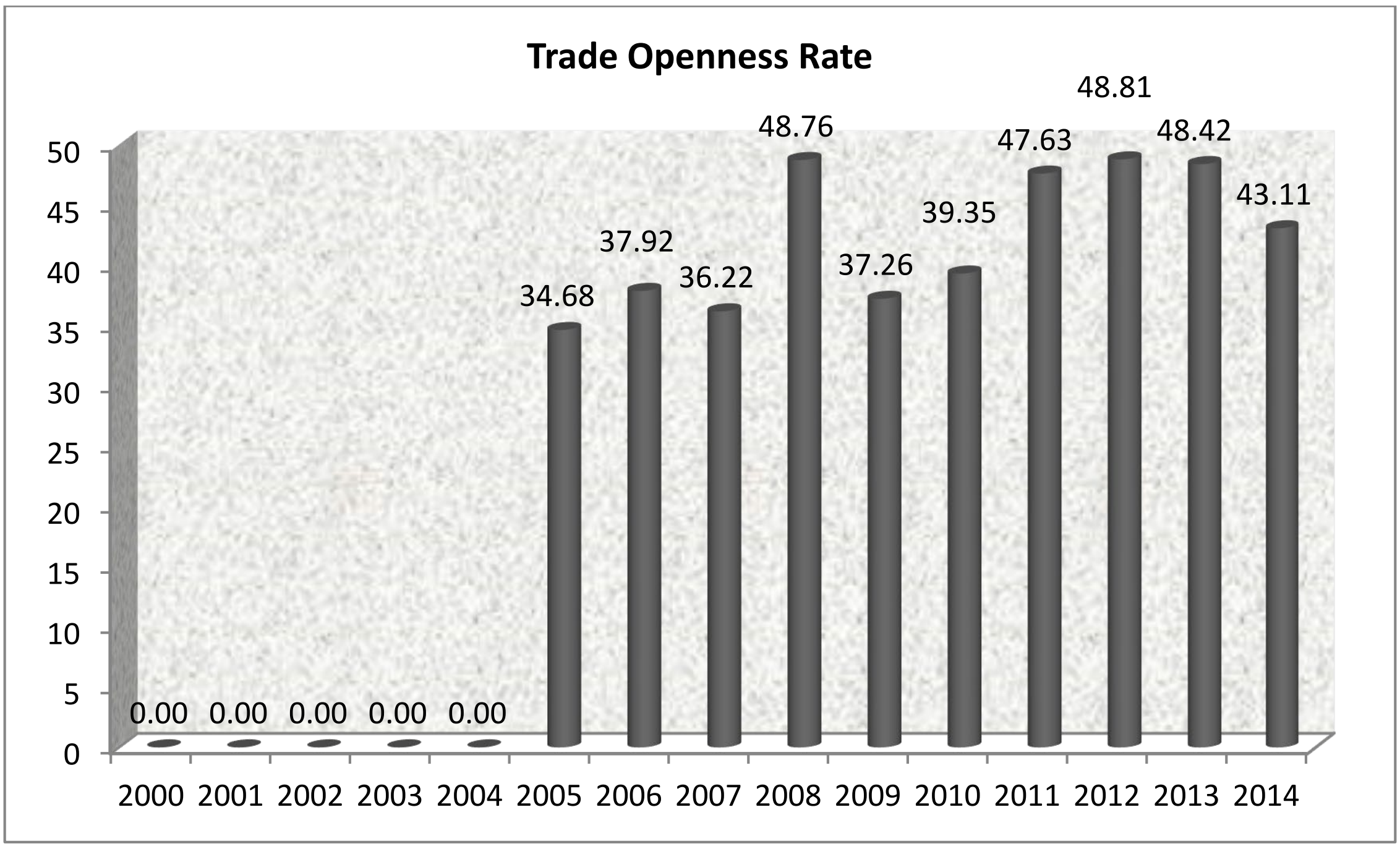

Figure 4: Rate of trade liberalization

Table 3. Year wise comparison of export-import of India with the world

| Year | Export of India | Import of India | Export of World | Import of World | % World Export | % World Import | % World Total Trade |
|---|---|---|---|---|---|---|---|
| 2000 | .. | .. | .. | : | : | : | : |
| 2001 | .. | .. | .. | : | : | : | : |
| 2002 | .. | .. | .. | : | : | : | : |
| 2003 | .. | .. | .. | : | : | : | : |
| 2004 | .. | .. | .. | .. | : | : | : |
| 2005 | 154582000000.00 | 134692000000.00 | 12885652214935.90 | 10056450421260.00 | 1.20 | 1.34 | 1.26 |
| 2006 | 193316000000.00 | 166572000000.00 | 14814722772433.60 | 11562144568439.30 | 1.31 | 1.44 | 1.36 |
| 2007 | 240082000000.00 | 208611000000.00 | 17292421915749.90 | 13353736991053.60 | 1.39 | 1.56 | 1.46 |
| 2008 | 305119000000.00 | 291740000000.00 | 19874074886082.80 | 15471808299657.60 | 1.54 | 1.89 | 1.69 |
| 2009 | 260847000000.00 | 247908000000.00 | 15896435311736.80 | 11876372433169.60 | 1.64 | 2.09 | 1.83 |

| Year | Export of India | Import of India | Export of World | Import of World | % World Export | % World Import | % World Total Trade |
|---|---|---|---|---|---|---|---|
| 2010 | 348035000000.00 | 324320000000.00 | 18868434334839.70 | 14462240189730.40 | 1.85 | 2.24 | 2.02 |
| 2011 | 446375000000.00 | 428021000000.00 | 22343724135264.90 | 17471688907154.90 | 2.00 | 2.45 | 2.20 |
| 2012 | 443845000000.00 | 450249000000.00 | 22685575445328.10 | 17624346977148.40 | 1.96 | 2.55 | 2.22 |
| 2013 | 467759000000.00 | 433760000000.00 | 23369674323705.10 | 17918138912797.10 | 2.00 | 2.42 | 2.18 |
| 2014 | 485843000000.00 | 405122000000.00 | 23631617161046.90 | 17945360114385.10 | 2.06 | 2.26 | 2.14 |

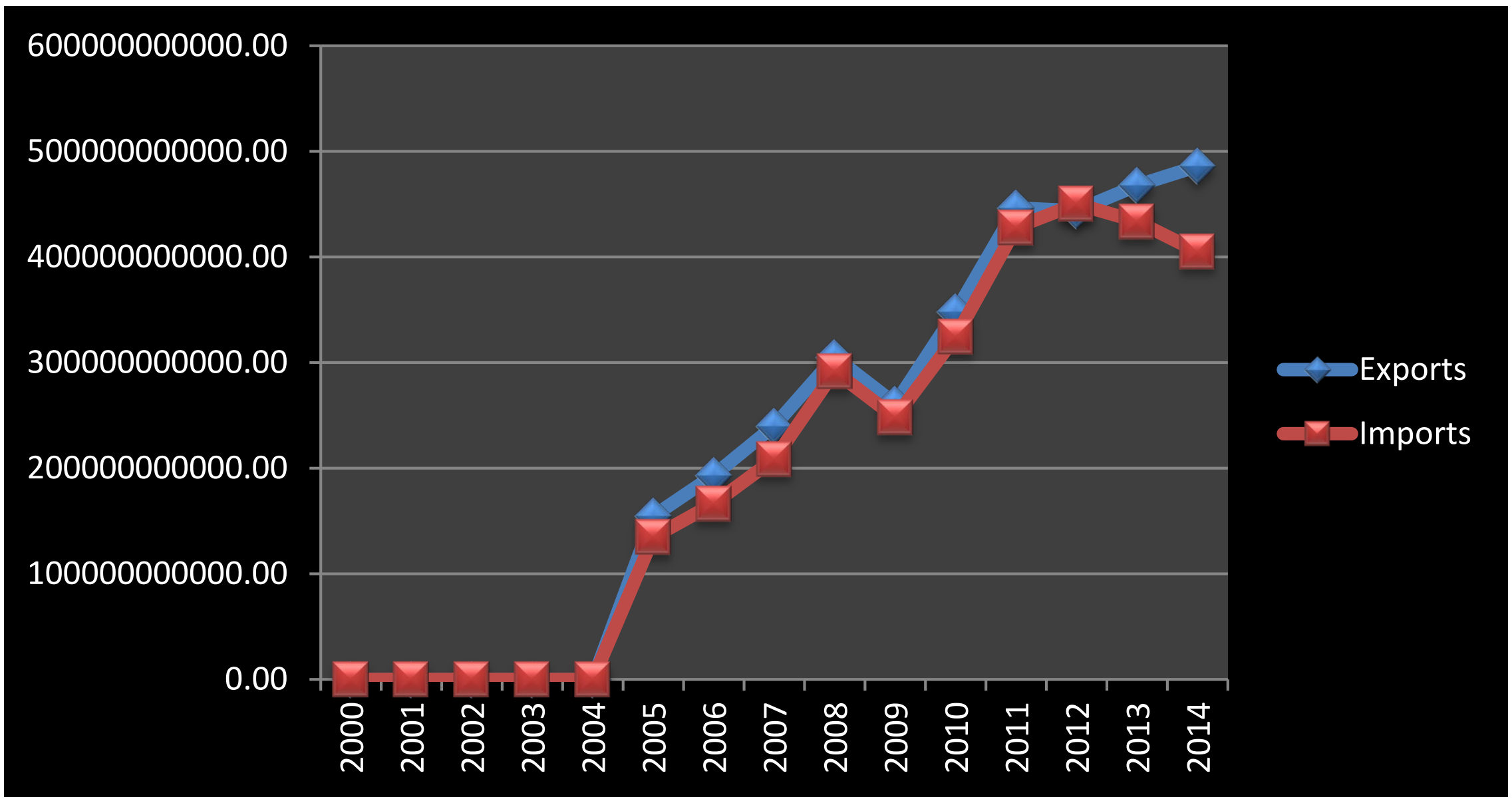

Figure 5: Exports and Imports of India

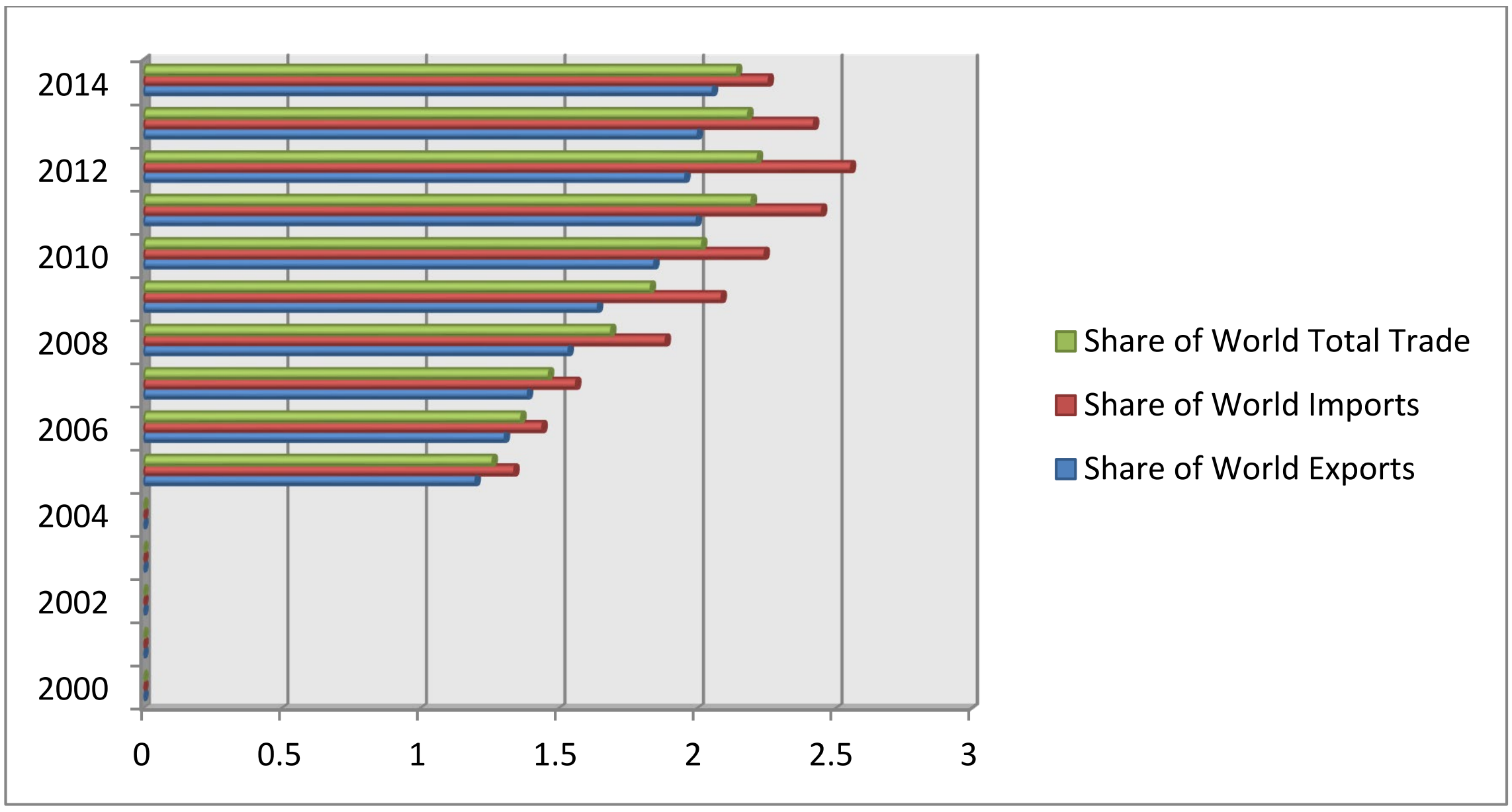

Figure 6: Participation of India in world trade